\documentclass{article}
\usepackage[nottoc]{tocbibind}
\usepackage{amsmath}
\usepackage{amssymb}
\usepackage{amsfonts}
\usepackage{amsthm}
\usepackage{enumerate}
\usepackage{authblk}

\usepackage[linesnumbered,lined,ruled,vlined,commentsnumbered]{algorithm2e}

\newtheorem{theorem}{Theorem}[section]

\newtheorem{proposition}[theorem]{Proposition}

\newtheorem{definition}[theorem]{Definition}
\newtheorem{assertion}[theorem]{Assertion}

\usepackage[colorlinks,linkcolor=blue,anchorcolor=blue,citecolor=blue,bookmarks = True]{hyperref}
\usepackage{longtable}
\usepackage{multirow}
\usepackage{float}
\usepackage{subfig}

\usepackage{booktabs}
\usepackage{graphics}
\usepackage{graphicx}
\usepackage[margin=25mm]{geometry}
\usepackage{color}
\usepackage{natbib}
\setcitestyle{authoryear,open={(},close={)}}
\usepackage{abstract}

\usepackage{graphicx}
\usepackage{verbatim}

\providecommand{\keywords}[1]
{
  \small    
  \textbf{\textit{Keywords:}} #1
}

\title{How Many Passengers Can We Serve with Ride-sharing?}
\author{Zeren Tan}
\date{August 2018}

\begin{document}

\maketitle
\begin{abstract}
Ride-sharing can reduce traffic congestion and thus reduce gas emissions and save travel time. However, transportation system with ride-sharing is currently inefficient due to low occupancy rate, high travel demand and some other factors. Existing literature did not consider ride-sharing with multi-request grouped in one trip. In our paper, we firstly proposed a graph-based algorithm that can obtain an approximation solution in polynomial time and then proposed an exact algorithm to solve this problem with maximizing the number of passenegers served in $O(1.2312^{|\mathcal{E}|})$ time.
\end{abstract}
\keywords{Ridesharing; Dynamic Matching; NP-Hard; Exact Algorithm}

\section{Introduction}
Traffic congestion, air pollution and many other societal and environmental concerns are rising with the rapidly increasing number of vehicles on the road. The cost of congestion in the United States are 1\% of its GDP, 5.5 billion hours of time lost to waiting in traffic congestion and 2.9 billion gallon of fuel wasted \citep{agatz2012optimization,alonso2017demand}. The greenhouse gas and toxic gas emissions and their negative consequences have not even been taken into account. Traffic congestion is caused in many ways, among which low car occupancy rate is a fatal one. According to European Environment Agency, the average number of passengers in a single vehicle ranges from 1.0 to 1.8. Similar result has also been found in the United States\citep{santos2011summary,agatz2012optimization}. Together with the high transportation demands, low vehicle occupancy rate leads to longer waiting time of a trip and sometimes longer distance travelled and finally costs the inefficiency of traffic system. 
\par
The huge costs brought by these negative effects of traffic congesion stimulate researchers, traffic operation agency as well as some companies to seek for a better solution. Big companies like Uber, Lyft and Didi have led a new transportation mode the so-called Mobility-on-demand (MoD) system, which improves the access to urban mobility with less waiting time and stress. MoD system is a more flexible and public transportation system that allows travelers to have more diverse demands. It is a user-centric approach which leverages emerging mobility services, integrated transit networks and operations, real-time data, connected travelers, cooperative Intelligent Transportation System (ITS) and incorporate shared-use and multimodal integration. 
\par
In ride-sharing, travelers with similar itineraries (i.e. similar departure time, similar arrival time, similar origin and destination) are grouped together. With ride-sharing, the driver involved in a trip can get more revenues without driving longer distance and passengers sharing a ride should pay less for this ride as they share their ride with others. The travel expenses are reallocated among ride-share participants but different passengers should benefit differently. Some participant may travel longer distance in order to pick up other participants and sometimes cannot reach his/her destination on time while other participants just share a ride with others without travel more. The way to determine how to reallocate the trips costs is an important problem that some researchers are working on. 
\par
Existing studies mainly focus on ride-sahring without pooling requests\citep{pavone2012robotic,spieser2014toward,zhang2016control,spieser2016shared}. But they did not consider servicing several travel request into a single trip. The best result we know is \citep{alonso2017demand}. They proposed a near-optimal solution that is tractable in practical application.

\section{Problem Statement}
\subsection{Assumptions}
There are several major assumptions related to the following discussion:
\begin{enumerate}
\item The time interval considered is sufficiently small so that there will not be any extra taxis arriving except the taxis we have observed at the beginning.
\item For each passenger, if the gain he get for sharing a ride is greater than the loss, which is acceptable, he is willing to take the ride.
\item A taxi driver is willing to give a ride as long as the distance is with his acceptable distance and he can get the mean value of the income he used to get regarding of similar time interval and similar travel distance.
\end{enumerate}
\begin{longtable}{rl}
        \caption{Notations}
            \label{tab:not}\\	
    \toprule
       Notation  & Description \\
       \endfirsthead
       Notation & Description \\
		\midrule
		\endhead
		\bottomrule \multicolumn{2}{r}{\textit{Continued on next page}} \\
		\endfoot
		\endlastfoot
       \midrule
        $\mathcal{P}$ & The set of passengers\\
        $N_i$ & The set of passengers who are \textit{close} (see Definition~\ref{def:1}) to passenger $i$\\
        $\mathcal{O}$ & The set of origins\\
        $\mathcal{D}$ & The set of destinations\\
        $\mathcal{DT}$ & The set of expected departure time\\
        $\mathcal{AT}$ & The set of latest arrival time\\
        $G_i$ & The group of passengers, which contains $i$, who are all \textit{close} to each other\\
        $M_i$ & The set of possible group rides for passenger $i$\\
        $Loss_i^m$ & The loss of passenger $i$ to join the group ride $m$\\
        $\mathcal{O}_f^m$ & The final departure location for group ride $m$\\
        $\mathcal{D}_f^m$ & The final destination for group ride $m$\\
        $\mathcal{DT}_f^m$ & The final departure time for group ride $m$\\
        $\mathcal{AT}_f^m$ & The final arrival time for group ride $m$\\
        $\mathcal{M}$ & The set of all possible trips\\
        $Dis_i^m$ & The discount passenger $i$ gain in ride $m$\\
        $Pay_i^m$ & The amount of money passenger $i$ pays the driver in group ride $m$\\
        $f_i$ & The amount of money passenger $i$ should pay the driver without sharing a ride\\
        $\delta$ & The constant threshold for distance in the definition of \textit{close}\\
        $t$ & The constant threshold for time in the definition of \textit{close}\\
        $\Delta _m$ & The average income for driver matched with group ride $m$ that he could get without ride-sharing\\
        $r_m$ & Passengers who take the ride $m$\\
        $c_i^m$ & An indicator variable, 1 if passenger $i$ take the group ride $m$; 0 otherwise\\
        $\zeta$ & The acceptable loss for each passenger when sharing a ride\\
        $\mathcal{B}$ & The set of available taxis\\
        $\overline{v}$ & The average driving velocity\\
        $z_b^m$ & An indicator variable, 1 if taxi $b$ is matched with group ride $m$; 0 otherwise\\
        $d_b^m$ & The distance taxi $b$ should travel in order to start trip $m$\\
        $\xi$ & The acceptable distance drivers are willing to travel in order to start a trip\\
\bottomrule
\end{longtable}

\begin{definition}[\textit{Close}]
For passengers $i,j \in \mathcal{P}$, they are \textbf{\textit{close}} to each other if and only if $||\mathcal{O}_i - \mathcal{O}_j||_M \leq \delta $,  $||\mathcal{D}_i - \mathcal{D}_j||_M \leq \delta$ and $|\mathcal{DT}_i- \mathcal{DT}_j| \leq t$, where $||\cdot||_M$ stands for the Manhattan distance.
\label{def:1}
\end{definition}
\par
We introduce this definition to ensure each passenger in a reasonable group of two or more passengers does not need to walk a lot and wait for too long in order to start a trip or reach his destination. If it is satisfied, we can simply ask them to meet at the geometric center of their requested origins and depart for the geometric center of their expected destinations. If it is not satisfied, a passenger may not want to share a ride with others and may violate.
\[
\begin{aligned}
  \max & \sum_{b\in\mathcal{B}}\sum_{m\in\mathcal{M}} |r_m|z_b^m 
\end{aligned}
\]

\begin{align}
    s.t. \sum_{m\in M_i} c_i^m \leq&  1 \quad \forall i\in \mathcal{P}  \label{eq:c1}\\
    c_i^m =  & c_j^m \quad \forall i,j \in r_m \quad \forall m\in\mathcal{M} \label{eq:c2}\\
    \sum_{b\in\mathcal{B}}z_b^m \leq& 1 \quad \forall m\in\mathcal{M} \label{eq:c3}\\
    \sum_{m\in\mathcal{M}}z_b^m \leq& 1 \quad \forall b\in \mathcal{B} \label{eq:c4}\\
    d_b^m \leq& \xi \label{eq:c5}\\
    Loss_i^m \leq & Gain_i^m\quad \forall i\in\mathcal{P}, m \in \mathcal{M} \label{eq:c6}\\
    \Delta_m  \leq & \sum_{i\in r_m} Pay_i^m \quad \forall m \in \mathcal{M} \label{eq:c7}\\
    Loss_i^m \leq& \zeta \quad \forall i\in\mathcal{P},m\in\mathcal{M} \label{eq:c8}\\
    c_i^m, z_b^m&\in\{0,1\}\quad \forall i\in \mathcal{P}, b\in \mathcal{B}, m\in\mathcal{M} \label{eq:c9}
\end{align}

where $Loss_i^m, Gain_i^m, Dis_i^m, f_i, \Delta_m$ are defined as follows:
\[
\begin{aligned}
Loss_i^m =& Function(||\mathcal{O}_i-\mathcal{O}_f^m||_M+||\mathcal{D}_i-\mathcal{D}_f^m||_M, |\mathcal{DT}_i-\mathcal{DT}_f^m|)\\
Gain_i^m = & Dis_i^m \cdot f_i\\
Dis_i^m = & Function(||\mathcal{O}_i-\mathcal{O}_f^m||_M+||\mathcal{D}_i-\mathcal{D}_f^m||_M, |\mathcal{DT}_i-\mathcal{DT}_f^m|, ||\mathcal{O}_i-\mathcal{D}_i^m||_M)\\
f_i = & Function(||\mathcal{O}_i-\mathcal{D}_i^m||_M)\\
\Delta_m = & Function(||\mathcal{O}_i-\mathcal{D}_i^m||_M,|\mathcal{DT}_f^m-\mathcal{AT}_f^m|)
\end{aligned}
\]
\par
The objective function is to maximize the total number passengers served by taxis. Constraint~\ref{eq:c1} suggests that each passenger can take at most one taxi and it is possible that some passengers may not have taxi to take. Constraint~\ref{eq:c2} ensures the validity of group ride $m$. Constraint~\ref{eq:c3} and Constraint~\ref{eq:c4} state that each taxi can be matched with at most one ride and each ride can be matched with at most one taxi. Constraint~\ref{eq:c5} and Constraint~\ref{eq:c7} guarantee that the solution is feasible for taxi drivers. Constraint~\ref{eq:c6} and Constraint~\ref{eq:c8} make sure that the rides grouped are feasible for passengers. Constraint~\ref{eq:c9} illustrates that the optimization problem is a 0-1 integer programming problem (IPP).

\section{A Graph-Based Algorithm}
\subsection{Group Ride and Initial Graph}
In order to group passengers, we introduce the definition of \textbf{stable pair}.
\begin{definition}[Stable Pair]
Two passengers $i,j\in\mathcal{P}$ are a \textbf{\textrm{stable pair}} if and only if they are \textbf{close} to each other and $Gain^m_i\geq Loss^m_i, Gain^m_j\geq Loss^m_j  \quad\forall m\in M_i\bigcap M_j$ .
\end{definition}
\begin{definition}[Stable Group]
A group of passengers is a \textrm{\textbf{stable group}} if and only if each two of them form a \textrm{stable pair}.
\end{definition}
From the definition of \textit{stable group}, it can be concluded that passengers can form a stable group ride are indifferent to their partners. We further assume that passengers in different stable groups cannot form any stable group. That is, we can divide passengers into several groups based on with regard to stable groups. Passengers in different stable group can never share a ride.  
\par
While this condition is somewhat restrictive, there are some practical applications
that meet this condition. For instance, in the international airport, passenger in Terminal 1 may not want to go to Terminal 2 or other terminals to take a taxi. Passenger in the same terminal can form a stable group while passengers in different terminal cannot. Passengers are naturally divided into several stable groups in this example.

\subsection{Graph Construction Algorithm}
\begin{figure}
    \centering
    \includegraphics[width = 4in]{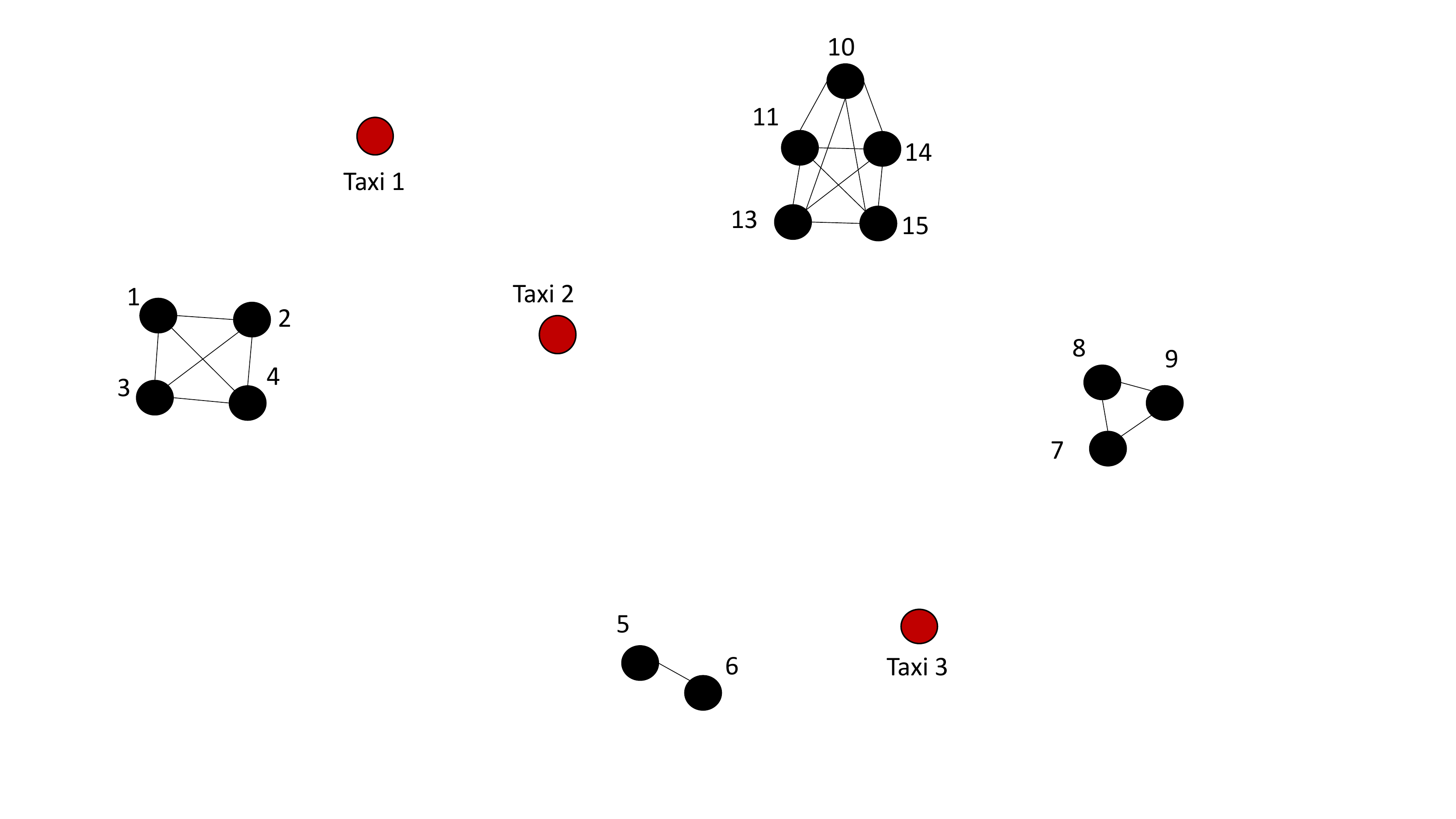}
    \caption{The initial graph}
    \label{fig:if}
\end{figure}

We input the stable groups of passengers and the set of taxis. In each group, passengers are all willing to share the same ride with other people. An example of the initial graph is presented in Figure~\ref{fig:if}. 
\begin{algorithm}[htbp]
\SetAlgoLined
\SetKwInOut{Input}{Input}\SetKwInOut{Output}{Output}
\caption{ConstructGraph($P, \mathcal{B}$)}
\label{alg:a5}
\BlankLine
Initialize $\mathcal{E}\leftarrow \emptyset$\;
\For{$p\in P$}{
	$\mathcal{E}\leftarrow \mathcal{E}\bigcup \{(p,t)\}$\;
	\For{$b \in \mathcal{B}$}{
		\If{$b$ is willing to give a ride to $p$}{
			$\mathcal{E}\leftarrow \mathcal{E}\bigcup\{(s,b), (b,p)\}$\;	
		}
	}
}
return $\mathcal{E}$
\end{algorithm}

\subsection{Algorithm Design}
In order to solve this problem, we choose to abstract this problem as a graph by the following steps.
\begin{itemize}
\item[Step 1] We abstract set of taxis and set of passengers as "small nodes".\label{step:1}
\item[Step 2] A stable group of passengers is abstracted as a "super node". \label{step:2}
\item[Step 3] There is an edge between the "super node" contains passengers and the taxi node if and only if the driver is willing to give the passengers a ride. \label{step:3}
\item[Step 4] We construct a graph $G=(V,E)$ with these nodes and add a source node and a sink node. The source node has edges with taxis nodes and the sink has edges with nodes contain passengers. \label{step:4}
\item[Step 5] Each edge $e\in E$ has a capacity $c_e$ which is defined as follows:
\[
c_e = \left\{
\begin{array}{cc}
\Big\lceil \frac{\displaystyle\textrm{number of passengers in } v}{\displaystyle4} \Big\rceil & e = (v, t) \\
1 & otherwise
\end{array}\right.
\]
\label{step:5}
\end{itemize}

\par
\begin{figure}
    \centering
    \includegraphics[width = 4in]{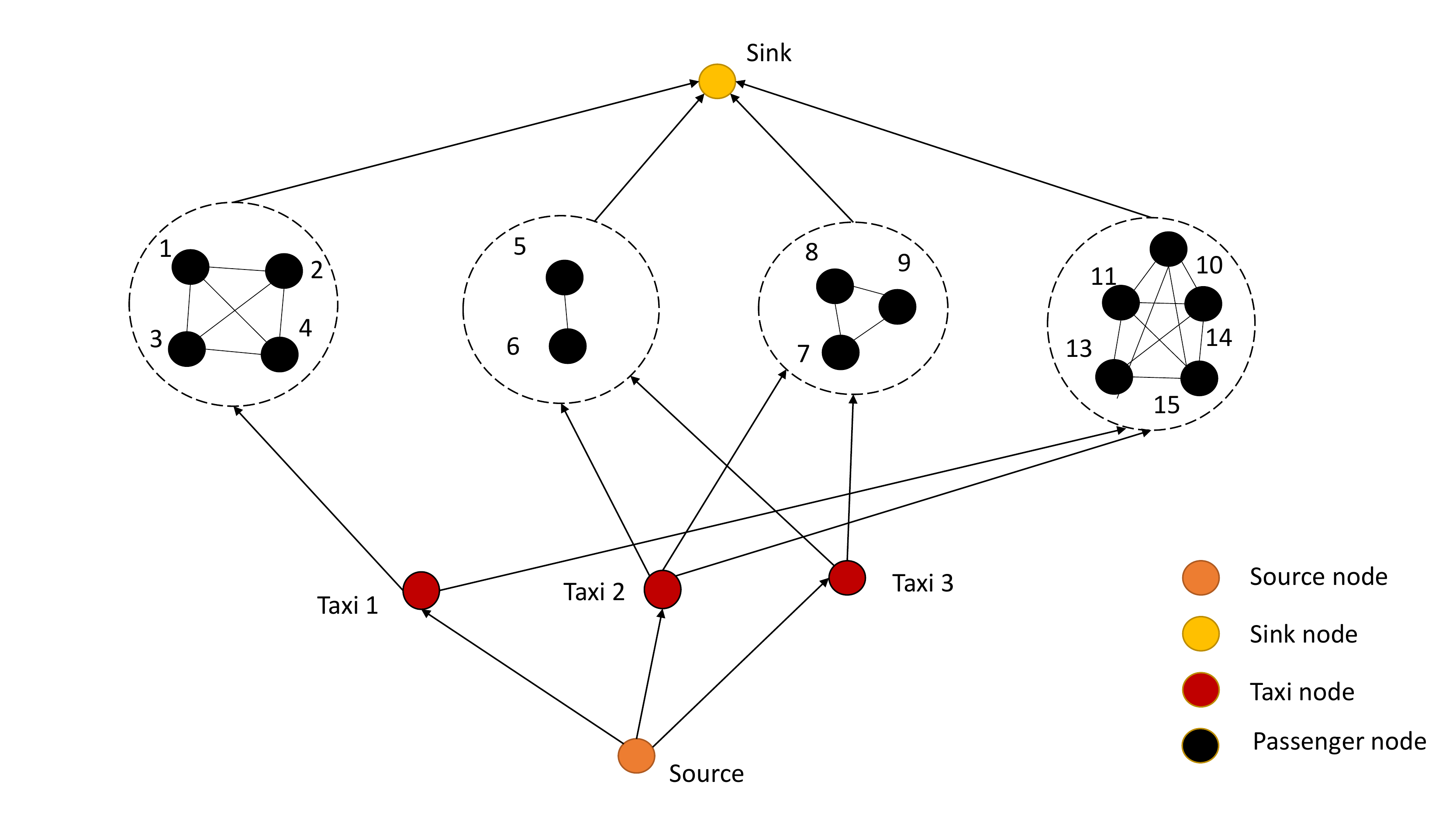}
    \caption{First Step of Graph Construction}
    \label{fig:CG}
\end{figure}
\begin{figure}
    \centering
    \includegraphics[width = 4in]{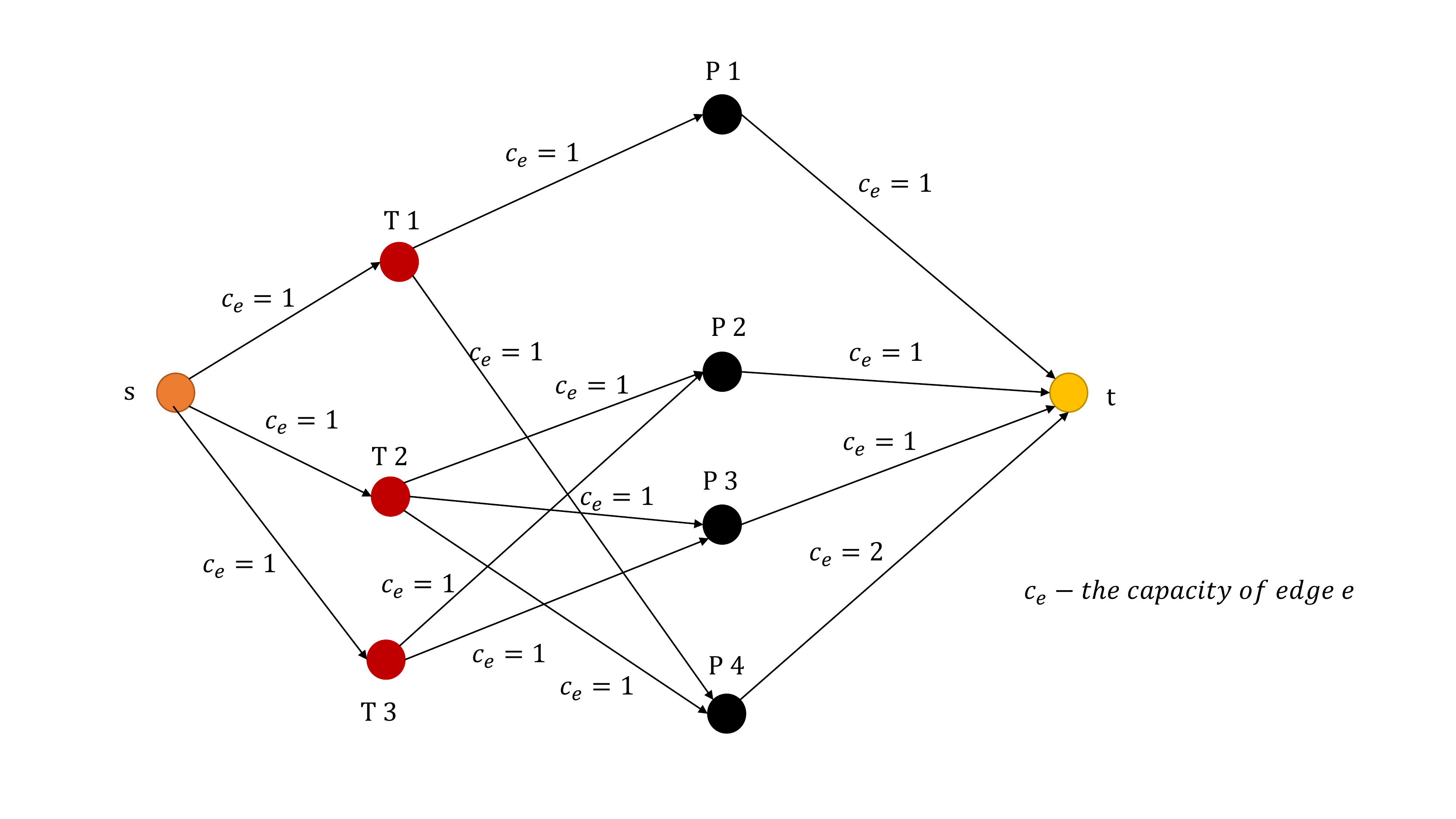}
    \caption{\textit{Super nodes} Abstraction}
    \label{fig:ag}
\end{figure}

Figure~\ref{fig:CG} and Figure~\ref{fig:ag} illustrate the abstraction process. 
\par
Everyone in the same passenger super node is indifferent to who are going to take the same taxi with him/her as long as they are in the same node. Thus, we are not concerned about the allocation of passengers to taxis. The main problem we are going to discuss is how to develop an algorithm that can determine the most effective distribution of taxis so that the system can serve the most passengers. That is, the algorithm we are going to develop can return how many taxis will be sent to each passenger super node. When taxis arrive, we can use constant time for each taxi to decide the way of assigning passengers. Because the amount of taxis arrive for each passenger super node $p$ is less than or equal to $\lceil \frac{|p|}{4}\rceil$ and the total number of taxis arrive is not greater than $|\mathcal{B}|$, the running time of assigning passengers is $O(|\mathcal{P}| + |\mathcal{B}|)$.
\begin{definition}[Maximum-Flow]
A flow network is a directed graph $G = (V,E)$ with a source node $s\in V$, a sink $t\in V$ and capacities along each edge. The amount of flow between two vertices is described by a mapping $f: V\times V\rightarrow \mathbb{R}$. The flow of a graph has the following properties:
\begin{align}\label{pro:1}
f_e \leq c_e
\end{align}
\begin{align}\label{pro:2}
f_{(u,v)} = -f_{(v,u)}\quad \forall u,v\in V
\end{align}
\begin{align}\label{pro:3}
\sum_{w\in V} f(v,w) = 0 \quad \forall v\in V-\{s, t\}
\end{align}
The total flow $|f|$ of a flow network is the amount of flow going out the source (or equivalently going to sink). Specifically, $|f| = \sum_{v\in V} f(s,v)$. The maximum-flow of a flow network is the largest possible flow for a given graph $G$.
\end{definition}
\par
We can show that the constraints stated above are all satisfied by the maximum-flow problem illustrated in Figure~\ref{fig:ag}.
\par
\begin{enumerate}[1.]
\item In the same passenger super node, origins and destinations of all passengers are close and when they are grouped together, their gain is greater than their loss. Thus, Constraint~\ref{eq:c6} and Constraint~\ref{eq:c8} are satisfied.
\item Taxis in the same node are willing to take the same rides. By Step~\ref{step:3}, Constraint~\ref{eq:c5} and Constraint~\ref{eq:c7} are satisfied.
\item  When we get a maximum-flow of $G$, there is a flow $f_e$ for each $e\in E$. We denote the set of taxi nodes as $T$ and the set of passenger nodes as $P$. For $p\in P$, denote the set of taxis that have edges with $p$ as $T_p$  If $e = (v,t)$ and $v\in P$, then $f_e$ indicates that passengers in node $v$ have $f_e$ taxis to take. And note that from the way we construct graph $G$, we can observe that $v\in P$ can only have one outgoing arc. The number of taxis available for $v$ is limited by Property~\ref{pro:1} and thus the value is $num = \min\{\sum_{u\in T_p}f_{(u,v)}, f_{(v,t)}\}$. When $v$ have $num$ taxis to choose, we assign some passengers to each taxi. In addition, it is important to note that both $f_e$ and the number of passengers assigned to a taxi can be zero. Property~\ref{pro:3} and Theorem~\ref{theo:1} ensure that a taxi is only assigned to one ride and a ride can only be matched with one taxi. Therefore, Constraint~\ref{eq:c1}$\sim$Constraint~\ref{eq:c4} are also satisfied.
\item  Theorem~\ref{theo:1} illustrate that there exist a integral maximum-flow since all capacities are defined as integers. Hence, Constraint~\ref{eq:c9} is satisfied too.
\begin{theorem}[Integral-flow Theorem]
Given a graph $G = (V, E)$, if $c_e\in \mathbb{Z}, \forall e\in E$, then there is a maximum flow in which all flows are integers.
\label{theo:1}
\end{theorem}
\end{enumerate}

\par
It is reasonable to assume that the taxi has taken as much passengers as it can in the node it is assigned.  We write as a proposition. 

\begin{proposition}\label{propo:1}
If $b\in \mathcal{B}$ is assigned to $p\in P$, then $b$ takes as much passengers as it can from $p$. 
\end{proposition}

The only problem of maximum-flow algorithm is that the solution can be optimal but sometimes it can also be very bad. The situation leads to bad solution is that sometimes a taxi is assigned to a passengers super node and there is only say 2 people take this taxi but in another passenger super node, which has an edge with the same taxi node, there are three or more people need this taxi. (Note that if the node that the waiting passengers are in does not have an edge with this kind of taxi node, the node contains these passengers must either have no edge with any taxis node or the solution is optimal. The first possibility means that these passengers cannot take any taxi. This is not what we are focusing on.)  Because of situations like this, the maximum-flow algorithm sometimes is not optimal for the taxis can take more passengers if assigned to another node. This problem often arises when there is not enough taxis and some passengers may not have taxi to take at that moment. We assert that if we can avoid this situation, we can get the optimal solution.
\begin{assertion}\label{ass:1}
If we can avoid the situation stated above, the solution found by maximum-flow algorithm is optimal.
\end{assertion}
\begin{proof}
The proof is obvious. If the situation does not happen in the solution found by our algorithm, it means that each taxi takes as many passengers as possible. Therefore, the total amount of passengers is as large as possible. That is to say, the solution serves the most passengers and thus is optimal.
\end{proof}
\par
We design an algorithm based on Ford-Fulkerson method. We add a score $g$ to each flow pass through node $p\in P$. The value of the score is defined as following: when there is a taxi $b_0$ assigned to node $v\in P$, the score this flow get from $v$ is $\min\{4n, |v|-\sum_{b\in T_p\backslash b, (b,v)\in E} f_{(b,v)}\}$, where $|v|$ is the number of passengers in $v$. The taxi $b$ will get the score $g_p^b$ if it is assigned to the node $p$. In order to get the optimal solution, we just augment 1 to flow value along a path $r$. The path $r$ has the property that when we augment 1 to flow value along it, we can get the most score compared to other augmenting paths. This strategy ensure that the aforementioned bad situation would not exist. The rigorous proof will be illustrated in \textit{Proof}~\ref{proof:1}. Th proposed algorithms are Algorithm~\ref{alg:a2} and Algorithm~\ref{alg:a3}.
\begin{algorithm}[htbp]
\SetAlgoLined
\SetKwInOut{Input}{Input}\SetKwInOut{Output}{Output}
\caption{OneAugment($f, g, c, r$)}
\label{alg:a6}
\BlankLine
\For{edge $e\in r$}{
	$u\leftarrow$ tail of $e$\;
	$v\leftarrow$ head of $e$\;
	\uIf{$e\in E$}{
		\If{$v\in P$}{
        	$g_v^e \leftarrow (g_v\mod 4)$\;
			$g\leftarrow g + g_v^e$\;
            $g_v\leftarrow g_v - g_v^e$\;
	}
		$f(e)\leftarrow f(e) + 1$\;
}\Else{
		\If{$u\in P$}{
        	$g\leftarrow g - g_u^{e^R}$\;
        }
		$f(e^R)\leftarrow f(e^R) - 1$\;
	}
}
return $f$ and $g$
\end{algorithm}

\begin{algorithm}[htbp]
\SetAlgoLined
\SetKwInOut{Input}{Input}\SetKwInOut{Output}{Output}
\caption{Modified-Ford-Fulkerson($G,s,t,c$)}
\label{alg:a7}
\BlankLine
\For{edge $e\in E$}{
	$f(e)\leftarrow 0$\;
}
\While{there exists some augmenting path $r$ in $G_f$}{
	$f, g\leftarrow$ \texttt{OneAugment($f,g,c,r$)}\;
	Update $G_f$\;
}
\While{there exist a passenger node $p_1$ for $b\in\mathcal{B}$ from which $b$ can get more scores}{
    $p_0\leftarrow$ the node $b$ is now sent to\;
    $g\leftarrow g + g_{p_1}^{(b,p_1)} - g_{p_0}^{(b,p_0)}$\;
   Update $G_f$\;
}
return $f$
\end{algorithm}
We assert that Algorithm~\ref{alg:a7} can terminate in finite time.
\begin{assertion}\label{ass:4}
Algorithm~\ref{alg:a7} will terminate in finite steps.
\end{assertion}
\begin{proof}\label{pro:4}
Because we only augment flow value by 1 per time and one upper bound for the maximum flow is $|\mathcal{B}|$, the first \texttt{while} loop will definitely stop in $O(|\mathcal{B}|)$ times. Since each time when the second \texttt{while} loop runs, the score for taxi $b$ will increase by 1, for each taxi, the \texttt{while} loop will only run 4 times. Therefore, the second \texttt{while} loop will run $O(|\mathcal{B}|)$ times.
\end{proof}

\begin{assertion}\label{ass:2}
The solution found by Algorithm~\ref{alg:a2} and Algorithm~\ref{alg:a3} is optimal.
\end{assertion}
\begin{proof}\label{proof:1}
The meaning of the score a taxi $b$ get is the number of passengers it takes and thus the total score $g$ represents the total number of passengers the assignment of taxis can serve. From Algorithm~\ref{alg:a6} and the augmenting step of Algorithm~\ref{alg:a7}, we can observe that $g$ and the score of a taxi $b$ are both non-decreasing with respect to the augmenting step.  
 In addition, note that each node $p\in P$ if ever matched, it would be always matched even though the set of taxis match with $p$ might change.
\begin{assertion}\label{ass:3}
After each augmenting step, the total score $g$ the flow obtain is non-decreasing.
\end{assertion}
\begin{proof}
We assume that the score before augmenting is $g_0$.
From Algorithm~\ref{alg:a6}, along an augmenting path $r$, we have three cases to discuss:
\begin{itemize}
\item[\textbf{\textit{Case 1}}] If there exist an edge $e_1\notin E, e_1^R\in E$, then there must exist $p_1\in P$ which is the tail of $e_1$ and $b_1\in\mathcal{B}$ which is the head of $e_1$. When augmenting, we assume that $e_2 = (b_2, p_1)$ is an edge of the augmenting path. Then, $g_{p_1}^{e_2} = g_{p_1}^{e_1^R}$. Assuming $e_3 = (b_1, p_2)$ is an edge of the augmenting path. We further assume that $p_2$ was not assigned any taxi before augmenting. There must exist this kind of node, otherwise, there will not have any augmenting path. Therefore, the total score $g  = g_0 + g_{p_2}^{e_3}>g_0$.
\item[\textbf{\textit{Case 2}}] If there does not exist $e\notin E, e_R\in E$ in augmenting path, then it is obvious that the flow value increase 1 and the score increase as well.
\item[\textbf{\textit{Case 3}}] If there is no augmenting path regarding of flow, then in the last \texttt{for} loop of Algorithm~\ref{alg:a7}, we can easily find that $g$ is non-decreasig.
\end{itemize}
To summarize, total score $g$ is non-decreasing.
\end{proof}

Assuming that the solution found by Algorithm~\ref{alg:a2} and Algorithm~\ref{alg:a3} is not optimal. Then, from Assertion~\ref{ass:1} and Proposition~\ref{propo:1}, we can conclude that the aforementioned situation must exist. That is, there exist one taxi $b_0$ that has edges with two passenger super nodes and $b_0$ is assigned to one node $node_1$ that $|q|$ passengers in $node_1$ take $b_0$ but in another node $node_2$ there is $|Q| > |q|$ passengers waiting for a taxi.
\par
If we think about why $b_0$ finally choose to take $q$, it turns out that when $b_0$ takes $q$, it can get more scores as how Algorithm~\ref{alg:a7} is implemented. The scores it has represent the number of passengers it takes.  Therefore, $|Q|\leq |q|$. This is a contradiction.
\end{proof}

\section{Running Time}
\subsection{Running Time of Algorithm~\ref{alg:a5}}
\par
In this section, we analyze the total running time of our algorithm.
\par
In Algorithm~\ref{alg:a5}, all the steps take $O(1)$. Therefore, the total running time is $O(|\mathcal{B}||P|)$.
\subsection{Running Time of Algorithm~\ref{alg:a6}}
\par
It is easy to observe that each path in $G_f$ has only $O(|\mathcal{B}|)$ edges. The step looking up if $v$ is in $P$ takes $O(1)$ on average. So the average running time for Algorithm~\ref{alg:a6} is $O(|\mathcal{B}|)$.

\subsection{Running Time of Algorithm~\ref{alg:a7}}
The running time for the initialization step is $O(|P|)$. Since $O(|\mathcal{B}|)$ for \texttt{OneAugment($f,g,c,r$)}, the steps in the first \texttt{while} loop take $O(|\mathcal{B}|)$. And this is the most expensive step in the first \texttt{while} loop. To compute the running time for the first \texttt{while} loop, we need to figure out the times this first \texttt{while} loop will run.
Since we only augment 1 flow value per time, the first \texttt{while} loop will run at most $|\mathcal{B}|$ times.  Therefore, the total running time is $O(|\mathcal{B}|^2)$. 
\par
The second \texttt{while} loop will run for at most $|\mathcal{B}|$ taxis and as stated in Proof~\ref{pro:4}, it will run at most 4 times for each taxi and $O(|P|)$ per time. Thus, the total running time for the second \texttt{while} loop is $O(|\mathcal{B}||P|)$.
\subsection{Total Running Time of Algorithms}
By summarizing the running time analysis of the above three subsections, we can conclude that the total running time is $O(|\mathcal{B}||P|+|\mathcal{B}|^2)$.

\section{Proof of NP-Hard}
We reduce our problem into a NP-Complete problem--Complete Coloring. In the language of graph theory, the decision problem of complete coloring can be phrased as following:
\par
\textit{
Given a graph $G = (V,E)$ and an integer $k$, find the answer that if there exist a partition of $V$ into $k$ or more disjoint subsets $V_1,V_2,\cdots,V_k$ such that each $V_i$ is an independent set of $G$ and for each pair of $V_i,V_j, V_i\bigcup V_j$ is not an independent set of $G$.
}
\par
\begin{proof}
Given an instance of Complete Coloring problem, we can create an instance of our problem by the following method. 
\par
\textit{Denote $P$ as the set of passengers and a passenger $p_i$ as a node. For each pair of passengers who cannot form a stable pair, we attach an edge $e$ with them. Assuming that for all vehicles $b\in\mathcal{B}$, it can take either of these passengers and the capacity of vehicle is large enough. }
\par
The mapping can be constructed as following: let each vertex $v_i$ in $G$ be vertex $p_i\in P$ and let independent set $V_i$ be vehicle $b_i\in B$. It can be observed that passengers that can form a stable trip can be collected as an independent set. So the nodes have the same color form a stable trip and are assigned into a vehicle. The set of passengers a vehicle contains is an independent set of the graph. It is obvious that if there is a solution for our problem, that is, if there exist an assignment of vehicles and passengers, the Complete coloring problem can be solved. If there is a partition of $G$ satisfies the aforementioned property, there is a assignment of vehicles and passengers.
\end{proof}

\section{A Branch and Bound Based Algorithm}
\subsection{Graph Construction Algorithm}
Algorithm~\ref{alg:a16}
\begin{algorithm}[htbp]
 \SetAlgoLined
\SetKwInOut{Input}{Input}\SetKwInOut{Output}{Output}
\caption{VTG($G$)}
\label{alg:a16}
\BlankLine
$\mathcal{T}\leftarrow \emptyset$;$\quad$ $\mathcal{E}\leftarrow \emptyset$\;

    \For{$i$ from 1 to $\nu$}{
        $\mathcal{T}_i\leftarrow \emptyset$\;
    }
    \For{$e = (b,v)\in G$}{
        $\mathcal{T}_1\leftarrow \mathcal{T}_1\bigcup \{v\}$\;
        $\mathcal{E}\leftarrow \mathcal{E}\bigcup \{(b,v)\}$\;
    }
    \For{$\forall v_1,v_2\in \mathcal{T}_1$ and $e = (v_1,v_2)\in G$}{
        $T \leftarrow\{v_1,v_2\}$\;
        $\mathcal{T}_2\leftarrow \mathcal{T}_2\bigcup T$\;
        $\mathcal{E}\leftarrow\mathcal{E}\bigcup \{(v_1, T),(v_2,T)\}$\;
        \If{$(b,v_1),(b,v_2)\in G$}{
            $\mathcal{E}\leftarrow\mathcal{E}\bigcup \{(b,T)\}$\;
        }
    }
    \For{$j$ from $3$ to $\nu$}{
        \For{$\forall T_1,T_2\in \mathcal{T}_{j-1}$ and $|T_1\bigcup T_2| = j$}{
            Denote $T_1\bigcup T_2 = \{v_1,v_2,\cdots,v_j\}$\;
            \If{$\forall k \in \{1,2,\cdots,j\}, \{v_1,v_2,\cdots,v_j\}\backslash v_k\in \mathcal{T}_{j-1}$}{
                $T\leftarrow T_1\bigcup T_2$\;
                \For{$i$ from 1 to $j$}{
                	$\mathcal{E}\leftarrow \mathcal{E}\bigcup \{(v_i,T)\}$\;
                }
                \If{$b$ can take trip $T$}{
                    $\mathcal{T}_j\leftarrow \mathcal{T}_j\bigcup T$\;
                    $\mathcal{E}\leftarrow \mathcal{E}\bigcup \{(b,T)\}$\;
                }
            }
        }
    }
    $\mathcal{T}\leftarrow (\bigcup_{i\in\{1,\cdots,\nu\}}\mathcal{T}_i)\bigcup\mathcal{T}$\;
\For{each vehicle $b\in\mathcal{B}$}{
    \For{$t\in \mathcal{T}$}{
        \If{$b$ is able to take trip $t$}{
            $\mathcal{E}\leftarrow\mathcal{E}\bigcup \{(b,t)\}$
        }
    }
}

\end{algorithm}
\subsection{Proof of Correctness}
\begin{figure}
    \centering
    \includegraphics[width = 5cm]{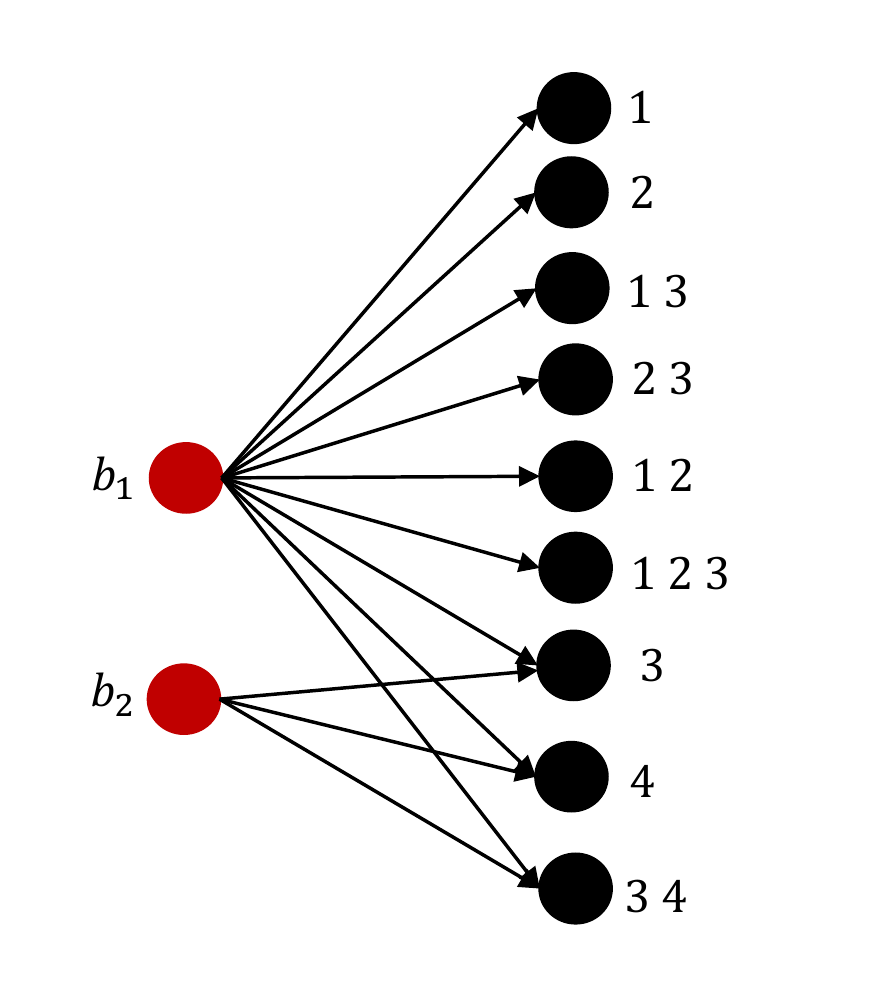}
    \caption{VTG}
    \label{fig:BB}
\end{figure}

\begin{algorithm}[htbp]
 \SetAlgoLined
\SetKwInOut{Input}{Input}\SetKwInOut{Output}{Output}
\caption{BranchBound($G$)}
\label{alg:a15}
\BlankLine
Find the set of connected components $\mathcal{C}$\;
\eIf{$|\mathcal{C}| > 1$}{
    $\mathcal{W}\leftarrow \emptyset$\;
    $\mathcal{S}\leftarrow \emptyset$\;
    \For{$c\in \mathcal{C}$}{
        $w_c,s_c\leftarrow \texttt{BranchBound}(G_c)$\;
        $\mathcal{W}\leftarrow\mathcal{W}+w_c $\;
        $\mathcal{S}\leftarrow\mathcal{S}\bigcup s_c $\;
    }
}{
    Enumerate all edges in $\mathcal{E}$ from $1$ to $n = |\mathcal{E}|$ if needed\;
    $e_1 = (u_1,v_1)\leftarrow$ the first edge of $\mathcal{E}$\;
    $G_1\leftarrow G$ removing $e_1$\;
    $G_2\leftarrow G$ removing $e_1,u_1,v_1$, all outgoing arcs of $u_1$ and all incoming arcs of $v$ and its neighbors\;
    \For{$v\in\mathcal{V}_2$}{
        \If{$u\cap u_1 != \emptyset$}{
            Remove $u$ and its incoming edges from $G_2$\;
        }
    }
    $\mathcal{W}_1,\mathcal{S}_1\leftarrow \texttt{BranchBound}(G_1)$;$\quad$ $\mathcal{W}_2,\mathcal{S}_2\leftarrow \texttt{BranchBound}(G_2)$\;
    \eIf{$\mathcal{W}_1 > \mathcal{W}_2$}{
        return $\mathcal{W}_1, \{s_1 = 0\}\bigcup\mathcal{S}_1$\;
    }{
        return $\mathcal{W}_2, \{s_1 = 1\}\bigcup\mathcal{S}_2$\;
    }
}
\end{algorithm}
\begin{assertion}
The algorithm can return the optimal solution for the original ILP.
\end{assertion}

\begin{proof}\makebox[4em]{} 
\begin{itemize}
   \item[Constraints] \begin{enumerate}[1.]
        \item When a vehicle is selected in the process, that is for one of its outgoing arc $e_i\in\mathcal{E}$,$s_i = 1$, it will not be selected in the upcoming process since all of its outgoing edges will be deleted once it is selected. This indicates that each vehicle is only matched with one trip.
        \item As long as for an edge $e_i$, $s_i = 1$, the incoming arcs of the matched trip and its neighbors will be removed from graph $G$. So a trip will be only matched with one vehicle.
        \item When deleting all the incoming arcs of a trip, the scanning process is going to detect which trip is incompatible with the selected trip and those trip will be remove from $G$ in the end of the scanning process. Hence, all passengers will be assigned to exactly one trip in the final solution.
    \end{enumerate}
    
    \item[Optimality] We prove this by contradiction.
    \par
    \makebox[2em]{}Suppose that there is an optimal solution of the ILP that can not be returned by the algorithm. Assuming that the selected edges of the optimal solution are $e_{n_1},e_{n_2},e_{n_3},\cdots,e_{n_k}$, this is $s_{n_i} = 1\quad \forall i\in\{1,\cdots,k\}$ should be returned by a correct algorithm. Then go back to see what the algorithm does. In the first step, $s_1 = 0$ or $1$ is determined by the algorithm. If $s_1$ is assigned as 0, only one arc will be deleted and the algorithm will see what $s_2$ should equal to. If $s_2 = 0$, then the algorithm will go on to see which value should be assigned to $s_3$. The process goes on. When the algorithm reach the point of deciding the value of $s_{n_1}$, either 0 or 1 is the possible choice. The algorithm will explore the two choices in the whole process.  When $s_{n_1}=1$ is going to be explored, the algorithm will go on to determine the value of $s_{n_1 +1},\cdots,s_{n_2-1}$. The assignment that $s_{k} = 0\quad\forall k\in\{n_1+1,\cdots,n_2-1\}$ is included in the search space of the algorithm. As long as the algorithm set all of $s_{n_1 +1},\cdots,s_{n_2-1}$ to be 0 when exploring, this assignment will be returned. And then the algorithm will determine the value of $s_{n_2}$. When it is going to exploring the solutions when $s_{n_2} = 1$, the process goes on. Until the algorithm is going to find what will happen if $s_{n_k} = 1$, the process ends since all the arcs are removed and this branch is done.
    \par
    \makebox[2em]{}From the above statement, it can be observed that the assignment $s_{n_1},s_{n_2},s_{n_3},\cdots,s_{n_k} = 1$ is included in the search space of the algorithm. Hence, if the solution is optimal, it will certainly be returned by the algorithm.
\end{itemize} 
\end{proof}

\subsection{An Example for Algorithm Implementation}
\begin{figure}
\centering
\includegraphics[width = 18cm]{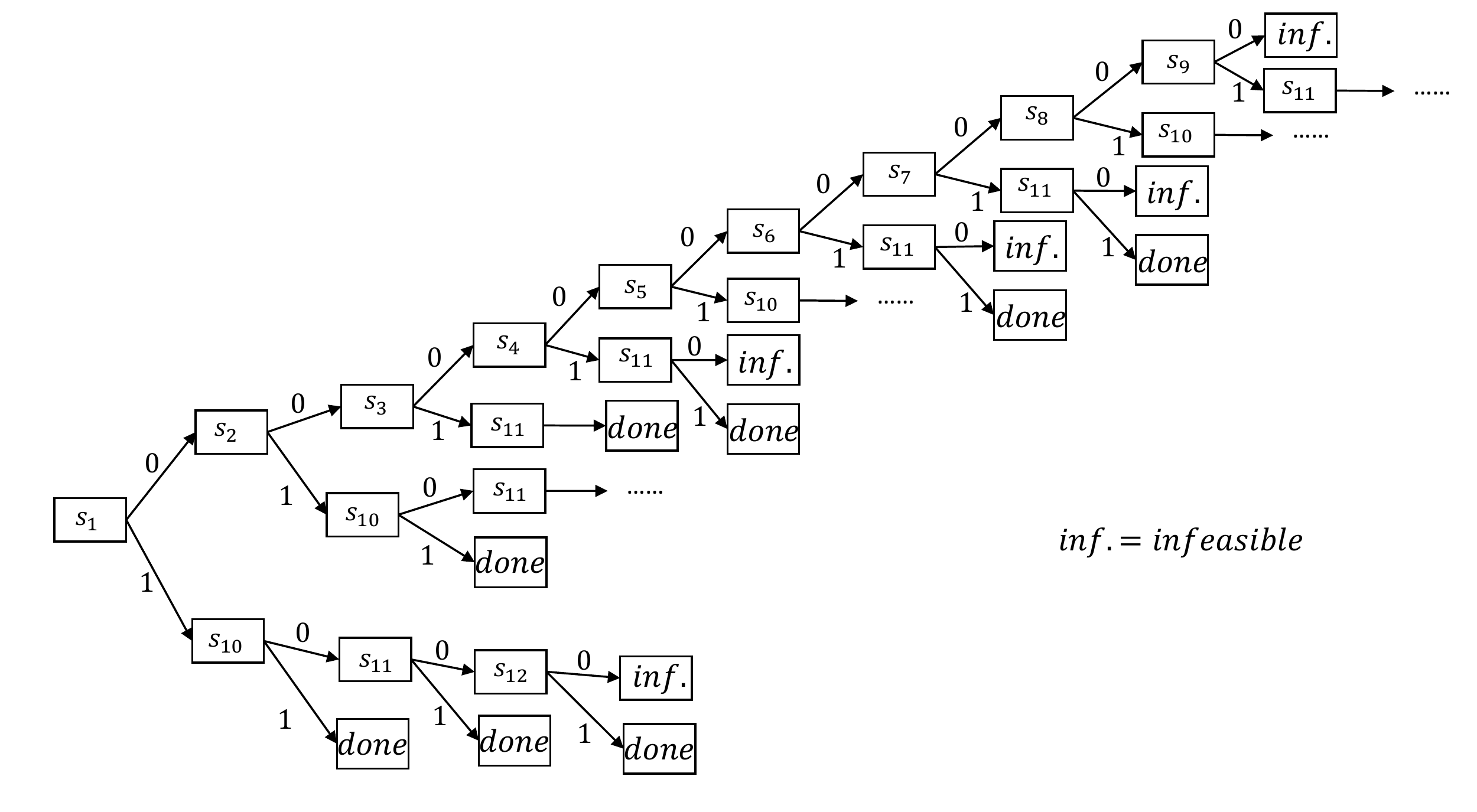}
\caption{Example}
\label{fig:9}
\end{figure}

Consider the Vehicle-Trip graph in Figure~\ref{fig:BB}, the steps for Algorithm~\ref{alg:a15} is shown in Figure~\ref{fig:9}.
\begin{itemize}
	\item[Step 1] Enumerate all edges as $1,2,\cdots,12$  from top to down.
	\item[Step 2] When $s_1 =1 $, vehicle $b_1$ is matched with trip one that contains passenger number one. Then arcs associated with $b_1$, node $1$ , node $1,2$, node $1,3$ and node $1,2,3$.
	\item[Step 3] Then it comes to the point of $e_{10}$. If $s_{10} = 1$, then it is done.
	\item[Step 4] If $s_{10} = 0$, then the algorithm will see the assignment of $s_{11}$. If $s_{11} = 1$ is assigned, then it is done.
	\item[Step 5] If $s_{11} = 0$, then the value of $s_{12}$ will be determined. If $s_{12} = 1$, then this path is finished.
	\item[Step 6] If $s_{12} = 1$, in this example, the solution is not optimal at all. Actually, it should be categorized as infeasible solution since $b_2$ is not matched with any trip while it really could.
	\item The steps go on like the above process. The algorithm will terminate when all possibilities are explored. The solution(s) with maximum passengers is returned as the result. For the final result, the comparison among these passenger-optimal solutions will be made in terms of the weight and the one with maximum weight will be returned in the end. This can be done in polynomial time.
\end{itemize}

\subsection{Running Time of Algorithm~\ref{alg:a15}}
From the algorithm, it can be implied that the running time of Algorithm~\ref{alg:a15} can be obtain by solving this following equation.
\begin{equation}\label{eq:17}
    T(|\mathcal{E^\prime}|) = T(|\mathcal{E^\prime}| - 1) + T(|\mathcal{E^\prime}| - d_b-\sum_{i = 0}^{d}d_{v_i} +d+1) + O(|\mathcal{B^\prime}|-1+|\mathcal{V^\prime}|-d-1)
\end{equation}

where $d_b,d_{v_i},d$ represent the outgoing, incoming degree and the number of neighbors of selected node, respectively, and $\mathcal{E^\prime},\mathcal{V^\prime},\mathcal{B^\prime}$ are the current edges, trip nodes, vehicles nodes, respectively.
\par
Next, the method of solving this equation is presented as following.
\begin{itemize}
    \item[Case 1] If $d = 0$, it means the trip only contains one passenger. then the problem can be reduced to a maximum weight bipartite matching problem. It can be solved in polynomial time by Hungarian method.
    \item[Case 2] If a vehicle only has edges with trips formed by a stable group, then we can simply match a vehicle with the node, which it connects with, contains the most passengers. This can be done in polynomial time too.
    \item[Case 3] If neither of the above cases exist, an approximation of $- d_b-\sum_{i = 0}^{d}d_{v_i} +d+1$ can be given. Since $d_b \geq 6$, $\sum_{i = 0}^{d}d_{v_i}\geq 2 (d+1)$ and $d\geq1$, then $d_b+\sum_{i = 0}^{d}d_{v_i} -d-1\geq 8$. In worst case, $d_b+\sum_{i = 0}^{d}d_{v_i} -d-1 = 8$.
    \par
    In this case, the original problem of solving Equation~\ref{eq:17} except the last term becomes solving the polynomial equation \begin{equation}
        x^8 = x^7 +1
    \end{equation}
    A real-value root of this equation is $x = 1.2321$. Since the last term of Equation~\ref{eq:17} is added only when the current edge the algorithm explores is assigned as 1, and in the final solution, the number of edges assigned as 1 is no more than the number of vehicles. In addition, the incidental term decreased when $s$ of one edge is assigned as 1.  Therefore, the incidental running time of scanning is $O(|\mathcal{B}|^2)$.  Hence, the running time of Algorithm~\ref{eq:17} is $O(1.2321^{|\mathcal{E}|}+|\mathcal{B}|^2)$ in terms of the worst case.
\end{itemize}

If the algorithm is expected to detect the infeasible solutions, then it needs more space to remember how many vehicles and stable groups left for each path. If the algorithm does not remove infeasible solutions and do not go on the detecting process , then it needs more space to remember the states of each edge.

\section{Experiments}
\subsection{Data}
The datasets from NYC TLC are used to evaluate the performance of the proposed algorithms. The datasets include the pickup and dropoff location and time of a recorded trip. We choose the data in a relatively small area of Manhattan and use trips detected in that place. To be specific, trips served in the rectangle area formed by () and (). Trips during 8:15-8:30 a.m. are extracted to verify the feasibility of our algorithms.

\section{Conclusion}
In this paper, we introduced a reactive anytime optimal method with scalable real-time performance for assigning passenger requests to a fleet of vehicles of varying capacity. We quantify experimentally the tradeoff between fleet size, capacity, waiting time, travel delay, and operational costs for low- and mediumcapacity vehicles, such as taxis or vans in a large-scale city dataset.Under the assumption of one person per ride, we show that 98\% of the taxi rides currently served by over 13,000 taxis could be served with just 3,000 taxis of capacity four. We observe that a vehicle capacity of two is sufficient for ride-sharing when a small trip delay of 2 min is imposed. If a maximum delay of 5 min or more (comparable to the time spent retrieving a car from parking) is allowed, higher-capacity vehicles (i) increase the service rate significantly, (ii) reduce the waiting time, and (iii) reduce the distance traveled by each vehicle. Our analysis shows that a ride-pooling service can provide a substantial improvement in urban transportation systems and that the system parameters such as vehicle capacity and fleet size depend on quality of service requirements and demand.

\bibliographystyle{plainnat}
\bibliography{Ridesharing.bib}
\end{document}